\documentclass[12pt]{article}

\usepackage{latexsym,amsmath,amssymb,amscd,epsfig}

\def\al{\alpha^{\prime}}
\def\a{& \hspace{-7pt}}
\newcommand\ortw[3]{\left[\frac{#1}{#3},\,
                          \frac{#2}{#3}
  \right](\tau)}
\newcommand\pportw[3]{\left[\frac{#1}{#3},\,
                          \frac{#2}{#3}
  \right]}
\def\@{@>>>\hspace{-1.3cm}\longleftarrow\hspace{.5cm}}

\def\bea{\begin{eqnarray}}
\def\eea{\end{eqnarray}}
\def\be{\begin{equation}}
\def\ee{\end{equation}}
\def\nn{\nonumber}
\def\bcd{\begin{CD}}
\def\ecd{\end{CD}}
\def\Z{{\bf Z}}
\def\z{{\mathbb Z}}
\def\F{{\mathcal F}}

\begin{document}

\thispagestyle{empty}

\begin{center}
\hfill SISSA-88/2002/EP \\

\begin{center}

\vspace{1.7cm}

{\LARGE\bf On the unfolding of the fundamental region in 
integrals of modular invariant amplitudes}

\end{center}

\vspace{1.4cm}

{\sc M.~Trapletti}\\

{\em ISAS-SISSA, Via Beirut 2-4, I-34013 Trieste, Italy} \\
{\em INFN, sez. di Trieste, Italy}
\vspace{.3cm}

\end{center}

\vspace{0.8cm}

\centerline{\bf Abstract}
\vspace{2 mm}
\begin{quote}\small
We study generic one-loop (string) amplitudes where an integration over the fundamental
region $\mathcal F$ of the modular group is needed. We show how 
the known lattice-reduction technique used to unfold $\mathcal F$ to
a more suitable region $S$ can be modified to
rearrange generic modular invariant amplitudes.
The main aim is to unfold $\mathcal F$ to the strip and, at the same 
time, to simplify the form of the integrand when it is a sum over a
finite number of terms, like in one-loop amplitudes for closed strings
compactified on orbifolds.
We give a general formula and a recipe to compute modular invariant amplitudes.
As an application of the technique we compute the one-loop vacuum
energy $\rho_n$ for a generic $\Z_n$ freely acting orbifold, generalizing
the result that this energy is less than zero and drives the system
to a tachyonic divergence, and that $\rho_n<\rho_m$ if $n>m$.

\end{quote}

\vfill

\newpage
\setcounter{equation}{0}

\section{Introduction}
\begin{center}
\begin{figure}[t]
\hspace{.5cm}
\includegraphics[scale=0.65]{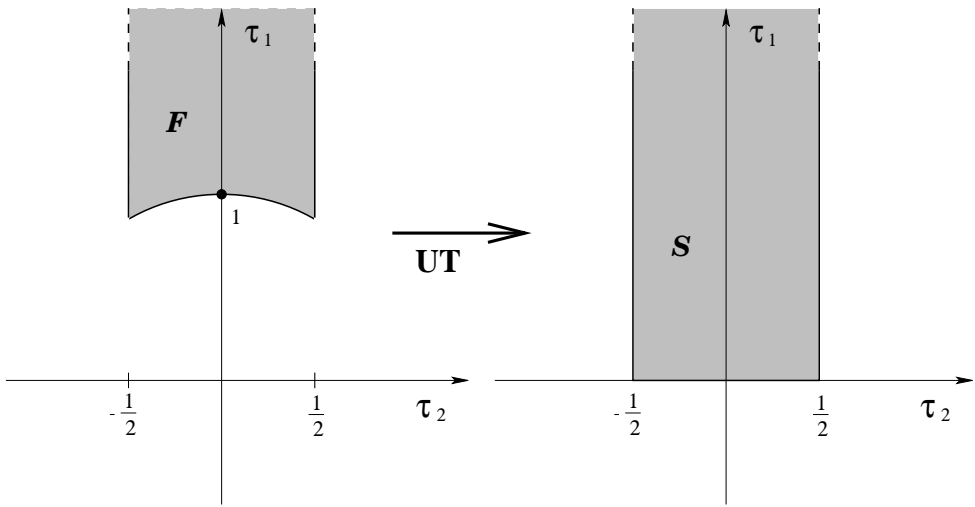}\vspace{-1.3cm}
\caption{\small The fundamental region $\F$, on the left and the strip $S$, 
on the right. The unfolding technique maps integrals over $\F$ into 
integrals over  $S$.}
\label{Fig1}
\end{figure}
\end{center}\vspace{-1cm}

Modular invariance is a key feature of one-loop amplitudes in closed string theory.
Usually it is gauged away by integrating the one loop amplitudes over a non trivial fundamental
region $\F$, described in the left side of fig.~\ref{Fig1}, instead then over the full 
complex plane $\mathbb{C}$.
This makes the integral difficult and usually it is necessary to introduce some technique to
unfold the region $\mathcal F$ to a simpler one \cite{Tan,Dix,MS,Mar,Ghil}.
Moreover, when considering a theory quantized on an orbifold, the typical one-loop amplitude is
given by the sum of a large finite number of terms; this
is due to the presence of twisted sectors in the
spectrum: in general an amplitude related to a $\Z_n$ orbifold contains  $n^2$ terms.

The main aim of this paper is to describe how the known lattice-reduction techniques
\cite{Tan,Dix,Mar} can
be modified and applied to amplitudes related to $\Z_n$ orbifold to unfold the fundamental
region $\mathcal F$ and, at the same time, to simplify the structure of the integrand itself.

The technique, initially studied for  $\Z_n$ orbifolds, is presented
in a fully general way and can be applied to every modular invariant amplitude integrated over the 
fundamental region $\mathcal F$.

We describe some possible applications of the technique, that turns out to be very useful
when considering one-loop closed string amplitudes to compute, for example, threshold corrections
to gauge couplings in heterotic string theory \cite{MS,Mar}, or mass corrections to closed string states or 
the one-loop vacuum energy \cite{Ghil,BD,root}. 
This last issue in particular is of great interest.
In a non-supersymmetric
string model where some dimension is compact, the compactification moduli are usually flat directions 
for the potential at tree-level. At one-loop this is no more true, and the dependence
of the vacuum energy on the compactification moduli is exactly the potential for them and from
it we can deduce the quantum stability of the model. Non-supersymmetric 
models, in particular when supersymmetry (SUSY) is broken
through the so called Scherk-Schwarz \cite{SS}  mechanism, have received great interest in the last years
\cite{ads1,adds,SeSc,sst,Cot} and computations about the quantum stability have been worked out in \cite{BD} 
for a $\Z_2$ case and in \cite{root} for two four-dimensional cases and a class of seven-dimensional
ones. The techniques developed in this paper can be directly applied to compute the one-loop vacuum energy 
in the most general case of orbifold; in particular, 
in a final section, we show in some detail how to work out the calculations 
whose results were given in \cite{root}.
We show the given results for a class of seven dimensional $\Z_n$ orbifolds 
with odd $n$ and we also describe the even-$n$ case.
We also show how the results can be interpreted from a field theoretical point of
view and we derive analytically the behavior of the vacuum energy $\rho_n$ at various $n$'s, 
obtaining the result $\rho_n<\rho_m$ when $n>m$ in the odd-$n$ case.
The even-$n$ case is also treated and we find that the behavior depends 
also on 
the volume of the two dimensions where $\Z_n$ acts as a rotation.

\section{Lattice reduction}
In this section we review the lattice reduction technique to unfold the fundamental
region $\mathcal F$ to the strip. Then we present the way in which we want to modify it
when the integrand is a finite sum of terms.
We start from the generic amplitude
\bea
\nn
A=\int_\F \frac{d^2\tau}{\tau_2^2} I(\tau),
\eea
where $\tau$ is a complex variable, $\mathcal F$ is the fundamental region and
and $I(\tau)$ is a modular invariant function, i.e. a function that is invariant under
the transformations
\bea
\nn
\tau\rightarrow\frac{a_1\tau+a_2}{a_3\tau+a_4},\,\,\,\,\,a_i\,\in \z,\,\,\,\,\,\, 
a_1a_4-a_2a_3=1.
\eea
These transformations can be represented via the group $PSl(2,\z)$ of the matrices
\bea\nn
\left(
\begin{array}{cc}
  a_1 & a_2 \\
  a_3 & a_4
\end{array}
\right),
\,\,\,\,\,a_i\,\in \z,\,\,\,\,\,\,a_1a_4-a_2a_3=1,
\eea
generated by
\bea\nn
S=\left(
\begin{array}{cc}
  0 & 1 \\
  -1 & 0
\end{array}
\right),\,\,\,\,\,
T=\left(
\begin{array}{cc}
  1 & 1 \\
  0 & 1
\end{array}
\right),
\eea
modded out a $\Z_2$ group acting as $a_i\rightarrow -a_i$.
The fundamental region is taken exactly to gauge away the modular invariance
in a one-loop string amplitude, where the integrating region was, originally, the full 
plane $\mathbb C$:
\bea\nn
\F=\frac{\mathbb C}{PSl(2,\z)}.
\eea

The integration is difficult due to the form of $\F$
but we can use modular invariance to unfold this region to a more suitable one.
In general this is done \cite{Tan} considering some special properties of 
the complex lattice 
\bea
\label{lator}
{\Lambda}(\tau)=\sum_{m,\,n\,\in\z}e^{-\frac{|m+n\tau|^2}{\tau_2}}.
\eea
For each $m$ and $n$ in the sum we can
compute the  maximum common divisor (MCD) $p=(m,n)$ and write (\ref{lator}) as
\bea\nn
{\Lambda}(\tau)=\sum_{p\in\z}
\hspace{-12pt}
\sum_{\scriptsize
  \begin{array}{c}
    c\in\mathbb N,\\ 
    d\,\in\z\,\left|\right.\,(c,d)=1
  \end{array}}
\hspace{-12pt}
e^{-\frac{p^2\cdot\left|c+d\tau\right|^2}{\tau_2}}\,;
\eea
now, as shown in \cite{Tan}, given any $c$ and $d$ such that $(c,d)=1$, there exists a class of
transformations 
\bea\nn
\left(
\begin{array}{cc}
  a_1 & a_2 \\
  c & d
\end{array}
\right)
\eea
in $PSl(2,\z)$ mapping 
\bea\nn
e^{-\frac{p^2\cdot\left|c+d\tau\right|^2}{\tau_2}}\rightarrow
e^{-\frac{p^2}{\tau_2}}.
\eea
These transformations are related each other by a $T$ transformation so that it is possible
to choose only a couple of $a_i$  such that the related transformation maps from $\mathcal F$ to $S$. 
So, given any $c,\,d$ in the sum, it is possible to choose one and only one transformation 
in $PSl(2,\z)/T$ mapping from $\mathcal F$ to $S$, and, in particular, the lattice can 
be written as the orbit under $PSl(2,\z)/T$ of the reduced lattice
\bea\nn
\hat{\Lambda}(\tau)=\sum_{m\in\z}e^{-\frac{p^2}{\tau_2}}.
\eea
Noting that the original lattice was modular invariant and so was the integration measure and
$I(\tau)$, the general amplitude $A$ can be written as
\bea
\label{general}
A=\hspace{-12pt}
\sum_{g\,\in\,PSl(2,\,\z)/T}\int_{\mathcal F}g\left\{
\frac{d^2\tau}{\tau_2^2} I(\tau)\times\frac{{\hat\Lambda(\tau)}}{\Lambda(\tau)}\right\}=
\sum_{g\,\in\,PSl(2,\,\z)/T}\int_{g\left\{\mathcal F\right\}}\hspace{0pt}
\frac{d^2\tau}{\tau_2^2} I(\tau)\times\frac{{\hat\Lambda(\tau)}}{\Lambda(\tau)}, 
\eea
where the right side of (\ref{general}) is obtained simply by a change of variables.
Now we can remember the definition of $\mathcal F$ and note that
\bea\nn
\sum_{g\,\in\,PSl(2,\,\mathbb Z)/T}\hspace{-18pt}g\left\{\mathcal{F}\right\}=\hspace{-18pt}
\sum_{g\,\in\,PSl(2,\,\mathbb Z)/T}\hspace{-8pt}g\left\{\frac{\mathbb C}{PSl(2,\z)}\right\}=
\frac{\mathbb C}{T}=S
\eea
where $S$, the orbit of $\F$ under the action of  $PSl(2,\z)/T$, is simply $\mathbb C/T$, i.e. the strip
$\tau_1\in\left[-1/2,1/2\right]$, $\tau_2>0$, described in the right side of fig.~\ref{Fig1}.
We obtain
\bea
\label{generalfinal}
A=\int_S\frac{d^2\tau}{\tau_2^2} I(\tau)\times\frac{{\hat\Lambda(\tau)}}{\Lambda(\tau)}.
\eea

This kind of ``unfolding technique''  (UT) is completely general
and can be applied independently of the form of $I(\tau)$, that is leaved unchanged. In some 
sense it closes the problem of unfolding the fundamental region $\F$.
It is also useful to note that the lattice at the denominator can be written in a simple way through
the elliptic theta functions defined for example in \cite{Pol}:
\bea\nn
\Lambda(\tau)=\sqrt{\tau_2}\left|\theta_2(2\tau)\right|^2.
\eea
As introduced this does not conclude the purpose of this paper: it is interesting 
to use modular invariance of $I(\tau)$ in a more subtle way to simplify the integration.
In general $I(\tau)$ is a sum of terms that are not modular invariant but that are in a finite 
dimensional representation, or multiplet, of the modular group
\bea\nn
I(\tau)=\sum_{i=1}^N I_i(\tau).
\eea
This happens, for example, when considering amplitudes arising from orbifold models.
If $N$ is large calculations can be really cumbersome, in this paper we show how 
it is possible to perform a particular UT to simplify it in a large class of cases.
 Essentially we split the UT in two steps.
In the first step we study the properties of the multiplet, and try to reduce it as the orbit of
a suitable element $I_{0}(\tau)$ under the action of a subset $G$ of the modular group,
\bea\nn
A^\prime=\int_\F\frac{d^2\tau}{\tau_2^2} \sum_{i=1}^N I_i(\tau)=
\int_\F\frac{d^2\tau}{\tau_2^2} \sum_{g\in G}g\left\{I_0(\tau)\right\}.
\eea
Then, noting that $I_0(\tau)$ is left invariant by a subgroup $\Gamma$ of the modular group, 
we look for a suitable lattice that is invariant under $\Gamma$ and 
that can be reduced through $\Gamma$ exactly as $\Lambda(\tau)$ is
reduced by $PSl(2,\z)/T$ to $\hat\Lambda(\tau)$
\bea\nn
1=\frac{\Lambda^\prime(\tau)}{\Lambda^\prime(\tau)}=\sum_{\gamma\,\in\,\Gamma}\gamma
\left\{
\frac{\hat\Lambda^\prime(\tau)}{\Lambda^\prime(\tau)}
\right\}.
\eea
So it is possible to write the integral as
\bea\nn
A^\prime=\int_\F\frac{d^2\tau}{\tau_2^2} \sum_{g\in G}g\left\{I_0(\tau)\right\}=
\int_\F\frac{d^2\tau}{\tau_2^2} \sum_{g\,\in\, G,\,\,\gamma\,\in\, \Gamma} 
g\circ \gamma\left\{I_0(\tau)\times\frac{\hat\Lambda^\prime(\tau)}{\Lambda^\prime(\tau)}\right\}.
\eea
It is possible to choose $\Gamma$ in such a way that 
\bea\nn
\sum_{g\,\in\,G}g\{\Gamma\}=\frac{PSl(2,\z)}{T}
\eea
so that
\bea
A^\prime=\int_S\frac{d^2\tau}{\tau_2^2} I_0(\tau)\times\frac{\hat\Lambda^\prime(\tau)}{\Lambda^\prime(\tau)}.
\eea

In the next section we describe these two steps for a large class of multiplets,
giving a systematic and general way to unfold this kind of integrals.

\section{Finite irreducible representation of the modular group and the unfolding technique}
In this section we explain how the ideas presented previously 
can be applied to various finite irreducible representations 
of the modular group. 
We  study the general properties of the multiplets, we show
the connection between the classification of the infinite dimensional subgroup of the modular group
and the possible multiplet. We describe how the unfolding
can be worked out for multiplets related to a large  class of subgroups and we see how this analysis
is independent of the details of the multiplets.

A finite-dimensional representation is completely defined when we give the net of identifications 
of its elements under the action of the modular group. 
As in the example given in the previous section, given any multiplet
$\{I_i(\tau),\,\,i = \,1,\dots,N\}$ there is an $N$ dimensional
subset $G$ of $PSl(2,\z)$, $G=\{g_i,\,\,i = \,1,\dots,N\}$ such that 
$I_i(\tau)=g_i I_0(\tau)$.
Since we are interested only in the modular properties of the multiplet all
the information we need are contained in $G$.
We have to note also that, given a multiplet, the classification made
with $G$ is not one to one: given $G$ mapping $I_0$ in all the multiplet
in general it does not map also $I_1$ in the multiplet, so that each multiplet
is related to more than one $G$.
Since we are mainly interested in multiplets where at least one element is
$T$ invariant we call $G$ the set of elements  that maps the
generic $T$-invariant term in all the others\footnote{As we will show
this $G$ is unique also when there are more than one $T$-invariant terms.}.
Furthermore given $G$ is defined uniquely also the subgroup $\Gamma$ 
such that given any $\gamma\,\in\,\Gamma$, $\gamma I_0(\tau)=I_0(\tau)$ and
$\sum_{g\in G}g\{\Gamma\}=PSl(2,\,\z)$.
We classify each multiplet using this last subgroup $\Gamma$.

We begin with the class of multiplets relative to $\Gamma_0^1[n]$, the subgroup of 
$PSl(2,\z)$ of the matrices of the form:
\bea\nn
\left(
\begin{array}{cc}
n a+1 & b\\
nc & n d+1
\end{array}
\right)\sim
\left(
\begin{array}{cc}
n a-1 & b\\
nc & n d-1
\end{array}
\right)
\eea
where the identification is due to the equivalence between a matrix $A$ and $-A$.
We present firstly the example $n=5$, then we  generalize the result to any prime
number $n$ and then to any $n\neq 4$. The $n=4$ case is treated separately.

As said we are interested only in the modular properties of the elements and we can describe completely
them, at least at this stage, giving their transformations under $T$ and $S$.
This means that the net of identifications is important, while the ``names'' 
we use for the elements is not. Nevertheless to simplify the notation we introduce
here the names used to identify the various terms arising from the
computation of the vacuum energy for a $\Z_n$ orbifold.
We define
\bea\nn
\ortw{i}{j}{N}=
Tr\Bigg|_i\left[\theta^j q^{L_0}\bar q^{\bar L_0}\right],\,\,\, q=e^{2 \pi i \tau}
\eea
where $\theta$ is the $\Z_n$ generator, $L_0$ and $\bar L_0$ are the
Virasoro generators
and the trace is performed over the light-cone degrees of
freedom of the $\theta^i$ twisted sector of closed type IIB string theory, 
i.e. the sector where the world-sheet fields are identified as
\bea\nn
\Phi(s+2\pi,t)=\theta^i \Phi(s,t).
\eea
The trace contains also the usual GSO projection and a sum over spin structures.

In the $n=5$ case we refer to the twelve dimensional multiplet described by the graph
\bea\nn
\bcd
\ortw{0}{1}{5}  \@  \ortw{1}{0}{5}  @.    \ortw{2}{0}{5} \@  \ortw{0}{2}{5}   \\
      @.             @VVV                       @VVV              @.        \\
      @.         \ortw{1}{1}{5}     @.    \ortw{2}{2}{5}  @.                  \\
      @.             @VVV                      @VVV               @.        \\   
      @.         \ortw{1}{2}{5}  \@    \ortw{2}{4}{5}  @.               \\
      @.             @VVV           @VVV              @.        \\
      @.         \ortw{1}{3}{5}  \@    \ortw{2}{1}{5}  @.               \\
      @.             @VVV            @VVV             @.        \\  
      @.         \ortw{1}{4}{5}  @.    \ortw{2}{3}{5}  @.
\ecd
\eea
where the vertical and horizontal arrows refers to $T$ and $S$ transformations respectively,
$S$ and $T$ being  the previously described generators of the modular group.
In this and in all the nets we usually omit the vertical lines connecting $\ortw{i}{n-i}{n}$ to $\ortw{i}{0}{n}$
and the horizontal line between $\ortw{i}{i}{n}$ and $\ortw{i}{n-i}{n}$.

As it is clear there are two special $T$ invariant terms that are also invariant under the action of $\Gamma_0^1[5]$,
while the other terms are invariant under a subgroup that comes directly from $\Gamma_0^1[5]$.
As an example $\ortw{1}{0}{5}$ and $\ortw{2}{0}{5}$ are invariant under the action of $S\Gamma_0^1[5]S$.
Since there is a complete symmetry in the system, the unfolding can be performed equivalently from one 
or the other of the two $T$ invariant terms and as we will see the symmetry between
the two sectors is present also in the final result.

We have described the multiplet and concluded the first step: we take 
$I_0(\tau)$ to be $\ortw{0}{1}{5}$.
Now we can introduce the lattice
\bea
\label{ventitre}
{\Lambda_1}^{(5)} (\tau)=\sum_{m,\,n\,\in \mathbb Z} e^{-\frac{\left|5 m+1+5 n\tau\right|^2}{25 \tau_2}}=
\sum_{m,\,n\,\in \mathbb Z} e^{-\frac{\left|5 m+4+5 n\tau\right|^2}{25 \tau_2}}.
\eea
It is invariant under $\Gamma_0^1[5]$ and 
we can also note that the $p=(5m+1,5n)$ 
can be  $1$, $2$, $3$, $4$ $mod$ $5$, so that we can write (\ref{ventitre}) as
\bea\nn
{\Lambda_1}^{(5)} (\tau)=
\hspace{-20pt}
\sum_{\scriptsize
\begin{array}{c}
m,\,c,\,d\in\mathbb Z \\
(5c+1,5d)=1
\end{array}
}\hspace{-20pt}
e^{-(5 m+1)^2\frac{\left|5 c+1+5 d\tau\right|^2}{25\tau_2}}+
\hspace{-20pt}
\sum_{\scriptsize
\begin{array}{c}
m,\,c,\,d\in\mathbb Z \\
(5c+3,5d)=1
\end{array}
}\hspace{-20pt}
e^{-(5 m+2)^2\frac{\left|5 c+3+5 d\tau\right|^2}{25\tau_2}}.
\eea
Now we can consider the orbit of $Exp(\tau_2^{-1})$ under the action of $\Gamma_0^1[5]/T$ and obtain
\bea\nn
\sum_{\gamma\,\in\,\Gamma_0^1[5]/T}\gamma\left\{ e^{-\frac{1}{25\tau_2}}\right\}=
\hspace{-20pt}
\sum_{\scriptsize
\begin{array}{c}
c,\,d\in\mathbb Z \\
(5c+1,5d)=1
\end{array}}\hspace{-20pt}
e^{-\frac{\left|5 c+1+5 d\tau\right|^2}{25\tau_2}}.
\eea

We also note that under a $ST^2ST^3S$ transformation 
the subgroup $\Gamma_0^1[5]/T$ is shifted in the
subset of all the transformations of the form
\bea\nn
\left(
\begin{array}{cc}
n a+3 & b\\
nc & n d+2
\end{array}
\right)\sim
\left(
\begin{array}{cc}
n a-3 & b\\
nc & n d-2
\end{array}
\right).
\eea
We call this subset $\Gamma_0^2[5]$, and it is simple to verify that the orbit of 
$Exp(\tau_2^{-1})$ under the action of $\Gamma_0^2[5]$ is
\bea\nn
\sum_{\gamma\,\in\,\Gamma_0^2[5]/T}\gamma\left\{ e^{-\frac{1}{25\tau_2}}\right\}=
\hspace{-20pt}
\sum_{\scriptsize
\begin{array}{c}
c,\,d\in\mathbb Z \\
(5c+2,5d)=1
\end{array}}\hspace{-20pt}
e^{-\frac{\left|5 c+2+5 d\tau\right|^2}{25\tau_2}}.
\eea
Now it is clear that the lattice ${\Lambda_1}^{(5)}(\tau)$ can be written as
\bea\nn
{\Lambda_1}^{(5)}(\tau)=
\sum_{\gamma\,\in\,\Gamma_0^1[5]/T}\gamma\left\{\hat\Lambda_1^{(5)}(\tau)+ST^2ST^3S
\left\{\hat\Lambda_2^{(5)}(\tau)\right\}\right\}
\eea
where
\bea\nn
\hat\Lambda_i^{(n)}(\tau)=\sum_{m\in\mathbb Z}
e^{-\frac{(nm+i)^2}{n^2\tau_2}}.
\eea
This allows us to write
\bea\nn
 A_5&=&
\int_{\mathcal F} \frac{d^2\tau}{\tau_2^2}\sum_{i,\,j}\ortw{i}{j}{5}=
\sum_{g\,\in\,G}
\int_{\mathcal F} g\left\{\frac{d^2\tau}{\tau_2^2}\ortw{0}{1}{5}\right\}=\nn\\&&\hspace{-12pt}
\sum_{\scriptsize\begin{array}{c}g\,\in\,G,\\ \nn \gamma\,\in\,\Gamma_0^1[5]/T\end{array}}\hspace{-12pt}
\int_{\mathcal F} g\circ\gamma 
\left\{\frac{d^2\tau}{\tau_2^2}\frac{\ortw{0}{1}{5}}{{\Lambda_1}^{(5)}(\tau)}\times
  \left[ 
    \hat\Lambda_1^{(5)}(\tau)+ST^2ST^3S\left\{\hat\Lambda_2^{(5)}(\tau)\right\}
  \right]
\right\}.
\eea
Now we note that $ST^2ST^3S$ maps exactly $\ortw{0}{1}{5}$ 
in $\ortw{0}{2}{5}$ and ${\Lambda_1}^{(5)}(\tau)$ in
${\Lambda_2}^{(5)}(\tau)$, where ${\Lambda_i}^{(5)}(\tau)$ 
is a simple extension of the previous definition:
\bea\nn
{\Lambda_i}^{(5)} (\tau)=\sum_{m,\,n\,\in \mathbb Z} e^{-\frac{\left|5 m+i+5 n\tau\right|^2}{25\tau_2}}=
\sum_{m,\,n\,\in \mathbb Z} e^{-\frac{\left|5 m-i+5 n\tau\right|^2}{25\tau_2}},
\eea
and, furthermore, we note that $\Gamma_0^1(5)$ commutes with  $ST^2ST^3S$ 
and $G\circ ST^2ST^3S\sim G$, so that,
\bea
A_5=\int_{\mathcal F}\frac{d^2\tau}{\tau_2^2} \sum_{i,\,j}\ortw{i}{j}{5}=
\int_{\mathcal S} \frac{d^2\tau}{\tau_2^2}
\sum_{i=1}^2\frac{\ortw{0}{i}{5}}{{\Lambda_i}^{(5)}(\tau)}\times
\sum_{m\in\mathbb Z} e^{-\frac{(5m+i)^2}{25\tau_2}}.
\eea
As anticipated the final form is symmetric in the $T$-invariant terms.

The symmetry of the result, that contains all the $T$ invariant terms,
is encoded in the fact that the invariance group is not the full subgroup 
$\Gamma_0[5]$ of the matrices 
\bea\nn
\left(
\begin{array}{cc}
 a & b\\
5c & d
\end{array}
\right)
\eea
as one can expect generalizing the results of \cite{MS}, but a subgroup of it, that can be mapped in the
full $\Gamma_0[5]$ by the set of transformations $\{I,ST^2ST^3S\}$, where $ST^2ST^3S$ maps exactly one
$T$-invariant term in the other. 

The extension to a general $\Gamma_0[n]$ for a generic prime number $n$ is straightforward.
The  multiplet  contains $(n^2-1)/2$ terms, joined in $(n-1)/2$ sectors  of the form 
\bea
\label{gensec}
\bcd
\ortw{0}{1}{n} @>S>>    \ortw{1}{0}{n} \\
          @.        @VVV     \\
          @.        \dots     \\
          @.        @VVV     \\
          @.      \ortw{1}{n-1}{n}
\ecd
\eea
with $(n+1)$ elements each.

There are $(n-1)/2$ $T$ invariant elements, and we can guess these elements are invariant under
the action of $\Gamma_0^1[n]$, while the other transformations in  $\Gamma_0[n]\setminus\Gamma_0^1[n]$ exchange
the various $T$ invariant terms.
So the first step of the unfolding is completed taking one of the $T$ invariant terms as
fundamental element and $\Gamma_0^1[n]$ as unfolding group.
The second step is a generalization of the calculations worked out in the $n=5$ case,
we have only to use $\Lambda_1^{(n)}$ and the final result is
\bea
\label{finalformula}
A_n=\int_{\mathcal F} \frac{d^2\tau}{\tau_2^2}\sum_{i,\,j}\ortw{i}{j}{n}=
\int_{\mathcal S} \frac{d^2\tau}{\tau_2^2}
\sum_{i=1}^{(n-1)/2}\frac{\ortw{0}{i}{n}}{{\Lambda_i}^{(n)}(\tau)}\times
\sum_{m\in\mathbb Z} e^{-\frac{(nm+i)^2}{n^2\tau_2}}.
\eea

Now we can study the case of multiplets relative to arbitrary $n$'s.
In this case the multiplet contains terms that are 
invariant under $\Gamma_0^1[n]$, i.e. there are parts of the multiplet of the form described in the graph 
(\ref{gensec}), but in general there are also different building blocks. This is not a problem and at this 
stage the full form of the multiplet is not indispensable.
 As a general recipe we can say that, given a certain multiplet, one 
sees if there exists a $T$ invariant term, then one studies the part of the multiplet related to this element
via $S$ and $T^aS$ transformations and finds the order of the invariance subgroup. Then the unfolding is 
completed by considering one of the invariant terms as generator
of the full multiplet and the subgroup $\Gamma_0^1[n]$ as base for the second step.
The presence of further $T$-invariant terms is encoded in  $\Gamma_0^1[n]$, exactly as 
explained in the $\Gamma_0^1[5]$ case.
Formula (\ref{finalformula}) can be applied with attention to the fact that if $n$ is not a prime number 
then the index $i$ does not run over $\{1,2,\dots,(n-1)/2\}$ but only on the numbers $p$ previously described.
For example in the $n=9$ case, of interest, for example, for the 
36-dimensional irreducible  multiplet arising when 
computing the free energy in a $\Z_9$ model, the index $i$ takes values $1,\,2,\,4$.
The only note is about the identification of the $T$-invariant terms of the multiplet analyzed
with the terms $\ortw{0}{i}{n}$ used in (\ref{finalformula}). This is done by choosing one 
$T$-invariant term and calling it  $\ortw{0}{1}{n}$, then one can derive the net of identifications
for the original multiplet and for  $\ortw{0}{1}{n}$ knowing that
\bea\nn
\bcd
\ortw{h}{k}{N} @>S>>    \ortw{k}{-h}{N} \\
       @VTVV              @.      \\
\ortw{h}{h+k}{N}  @.\\   
\ecd
\eea
and
$
\ortw{h}{k}{N}\sim\ortw{-h}{-k}{N}\sim\ortw{nN-h}{pN-k}{N},\,\,\forall\,\,n,\,p\,\in\mathbb N.
$
At this point the identification is easily done by comparing the two nets.

As an example we present the twelve-dimensional multiplet arising from a $\Z_6$ orbifold.
The full multiplet contains 36 elements, but is not irreducible, it is given by the sum of the usual
trivial one-dimensional multiplet,
a three-dimensional multiplet related to $\Gamma_0^1[2]$, two equivalent four-dimensional multiplets
related to  $\Gamma_0^1[3]$ and two equivalent twelve-dimensional multiplets related to $\Gamma_0^1[6]$:
${\bf 36}={\bf 1}\oplus{\bf 3}\oplus 2\times {\bf 4}\oplus 2\times {\bf 12}$. This latter multiplet has
the form:
\bea\nn
\bcd
\ortw{0}{1}{6} \@    \ortw{1}{0}{6} @.                  @.                \\
          @.        @VVV               @.                  @.      \\
          @.      \ortw{1}{1}{6} @.                  @.                \\
          @.        @VVV               @.                  @.      \\
          @.      \ortw{1}{2}{6} @.\longleftarrow\hspace{-9pt}
                                  \longrightarrow @.    
                                                        \ortw{2}{5}{6} \\
          @.        @VVV               @.                @VVV     \\
          @.      \ortw{1}{3}{6} \@   \ortw{3}{5}{6}    @.    \ortw{2}{1}{6} \\
          @.        @VVV              @VVV              @VVV     \\
          @.      \ortw{1}{4}{6} @.   \ortw{3}{2}{6}   \@    \ortw{2}{3}{6} \\
\ecd
\eea
where an horizontal line links $\ortw{1}{4}{6}$ to $\ortw{2}{1}{6}$. The multiplet contains the only $T$
invariant term $\ortw{0}{1}{6}$, as we expect from the fact that $1$ is the only integer satisfying 
the condition of being at the same time prime with $6$ and less than $3$.

The integral is:
\bea\nn
A_6=\int_{\mathcal S} \frac{d^2\tau}{\tau_2^2}
\frac{\ortw{0}{1}{6}}{\sum_{m,\,n\,\in\,\mathbb Z} e^{-\frac{\left|6m+1+6n\tau\right|^2}{36\tau_2} }}\times
\sum_{m\in\mathbb Z} e^{-\frac{(6m+1)^2}{36\tau_2}}.
\eea

This concludes the analysis of the multiplets with $T$-invariant terms, that can be summarized as follows.
For a given $\Gamma_0^1[n]$ multiplet the unfolded integral equals
\bea
\label{finalformula2}
A_n=\int_{\mathcal F}\frac{d^2\tau}{\tau_2^2} \sum_{i,\,j}\ortw{i}{j}{n}=
\int_{\mathcal S} \frac{d^2\tau}{\tau_2^2}
\sum_{i=1}\frac{\ortw{0}{i}{n}}{\Lambda_i^{(n)}(\tau)}\times
\hat\Lambda_i^{(n)}(\tau)
\eea
where $i$ is in the set of numbers such that $(i,n)=1$ and  $i<(n-1)/2$. 

This is not the only one class of multiplets. From $\Z_n\times \Z_n$ orbifolds one can see that there are 
multiplets where all the term are equivalent and where the invariance subgroup is not based on $T$ and $ST^nS$
but on $T^n$ and  $ST^nS$, the relevant subgroup being $\tilde\Gamma[n]$ made of matrices of the form:
\bea\nn
\left(
\begin{array}{cc}
n a+1 & n b\\
nc & n d+1
\end{array}
\right).
\eea

Due to the fact that is more difficult to take in account $\tilde\Gamma[n]/T$ \footnote{This is due to 
the fact that $T\circ\tilde\Gamma\notin\tilde\Gamma$}, or, equivalently, that all the terms in the 
multiplet are equivalent we deduce that the UT useful in this case is the general formula 
(\ref{generalfinal}) described in the second section.

\subsection{An exception: $\Gamma_0[4]$}
The $\Gamma_0[4]$ case represents an exception to the recipe previously given. It is the only exception
suggested by orbifold models, so, if we  conjecture the one to one correspondence 
multiplet/invariance group and
note that from orbifold models we have regular multiplets for all  $\Gamma_0[n]$ $n\neq 4$, than we can 
guess that all the $n\neq4$ cases are regular.

The $\Gamma_0[4]$ case is exceptional because the multiplet
\bea\nn
\bcd
\ortw{0}{1}{4} \@    \ortw{1}{0}{4}  @.       \\
          @.        @VVV                  @.  \\
          @.      \ortw{1}{1}{4}     @.       \\
          @.        @VVV                  @.  \\
          @.      \ortw{1}{2}{4}     \@ \ortw{2}{1}{4}       \\
          @.         @VVV                 @.  \\
          @.      \ortw{1}{3}{4}     @.
\ecd
\eea
contains two $T$-invariant terms even though only 1 respects the conditions $(i,4)=1$, $i<2$.
The unfolding is written in the same way as before, but now starting from one term we do not obtain the 
second one. This is due to the fact that the contributions obtained starting from one or the other 
term are equal, as we will see in the example in the next section.
We can conclude that, even though the multiplet is exceptional, the unfolding is not and formula 
(\ref{finalformula2}) can be applied as in the other cases.

\subsection{Two examples}

The $n=2$  and $n=3$ cases have already been taken into consideration in \cite{MS,Ghil} and in
\cite{MS} respectively.
These are two examples of how the technique can be applied efficiently in various cases.

In particular in \cite{MS} the authors computed the threshold corrections to gauge couplings 
in an orbifold compactification of heterotic string.
Following \cite{Dix} they started from a one-loop amplitude, clearly modular invariant, that,
due to the presence of twisted sectors and projections, is made of a finite sum of elements.
In particular they considered the case of the three and four-dimensional multiplet
\bea\nn
\bcd
\ortw{0}{1}{2} \@    \ortw{1}{0}{2}\\
          @.        @VVV\\
          @.      \ortw{1}{1}{2}
\ecd
\eea
 and 
\bea\nn
\bcd
\ortw{0}{1}{3} \@    \ortw{1}{0}{3}  \\
          @.        @VVV     \\
          @.      \ortw{1}{1}{3} \\
          @.        @VVV     \\
          @.      \ortw{1}{2}{3}
\ecd
\eea
The first step of the unfolding was performed as described here, while the second step
was treated in a modified version to better fit the form of the terms they took into account.

In \cite{Ghil}, instead, the vacuum energy for an orbifold of type IIB string theory
was computed.
In particular the authors performed the unfolding in the $n=2$ case in the same way here presented.
The result is essentially that given in (\ref{exc}). In the next section we show how the computation
can be extended to a generic orbifold through the UT shown.

\section{The vacuum energy for a class of seven-dimensional orbifold models}
In this section  we compute the one-loop vacuum energy for
a class of models where
type IIB string theory is compactified over
$\mathbb R^7\times (\mathbb C\times S/\Z_N)$.
The group $\Z_N$ acts as an order $N$ rotation on $\mathbb C$
and an order $N$ translation around the compact circle $S$, taken to be of radius
$R$. 
This realizes a SUSY breaking through the so-called Scherk-Schwarz mechanism \cite{SS}, whose 
application in string theory was studied in \cite{ads1,adds,SeSc,sst,Cot}.
We consider firstly the odd-$N$ case, then we extend the result to the even-$N$ one.

As seen in the appendix, independently of the details of the action 
of $\Z_N$ the full amplitude contains 
$N^2$ elements named $\ortw{i}{j}{N}$,
 usually in a reducible multiplet: 
\bea\nn
Z^{(9)}=\frac{1}{N}\int_{\mathcal F}\frac{d^2\tau}{2\tau_2^2}\sum_{i,\,j\,=0}^{N} \ortw{i}{j}{N}
\eea
where the $\ortw{i}{j}{N}$ are defined in the previous section and in the appendix.

The reduction of the multiplet is quite easy.
First of all the term $\ortw{0}{0}{N}$ is a singlet, 
and is SUSY, so its contribution is zero. Then, due to the
periodicity properties, the $(N^2-1)$-dimensional multiplet 
obtained is reduced into two equivalent $(N^2-1)/2$-dimensional multiplets.
We consider one of them. It contains the $(N-1)/2$ $T$-invariant terms 
$\{\ortw{0}{1}{N},\dots,\ortw{0}{(N-1)/2}{N}\}$.
These terms are the key ones, being, with the exception of the $\Z_4$ orbifold, the only $T$-invariant terms
in all the multiplet.
Now the reduction is done by considering the net for each of these multiplets, or 
better, some relevant features of the multiplet itself.
We consider the $\Z_9$ example first, then we see how the final form can be extracted 
in the general case without any further analysis.
In the $\Z_9$ case we have the term $\ortw{0}{1}{9}$ that is left invariant by the action of $\Gamma_0^1[9]$. 
The full multiplet, so,
contains also two other $T$-invariant terms, because the number of $p$ such 
that $(p,9)=1$ and $p<9/2$ is three:
$p=1,2,4$. Clearly these terms  are $\ortw{0}{2}{9}$ and $\ortw{0}{4}{9}$, while the term $\ortw{0}{3}{9}$ 
is in
an independent irreducible multiplet, based on $\Gamma_0^1[3]$.
The reduction is concluded, now we can apply directly formula (\ref{finalformula2}), since the
identification between the elements of our multiplets and those used in the previous section 
is trivial.
\bea
Z^{(9)}\a=\a2\int_{\mathcal S} \frac{d^2\tau}{\tau_2^2}
\sum_{i=1,\,2,\,4}\frac{\ortw{0}{i}{9}}{{\Lambda_i}^{(9)}(\tau)}\times
\sum_{m\in\mathbb Z} e^{-\frac{(9m+i)^2}{9^2\tau_2}}+\nn\\ \a\a
2\int_{\mathcal S}  \frac{d^2\tau}{\tau_2^2}
\frac{\ortw{0}{3}{9}}{{\Lambda_1}^{(3)}(\tau)}\times
\sum_{m\in\mathbb Z} e^{-\frac{(3m+1)^2}{3^2\tau_2}}
\label{zetanove}
\eea
where the overall factor of $2$ takes in account the terms $\ortw{0}{p}{9}$ with $p>4$.
Since 
\bea\nn
\sum_{m\in\mathbb Z}e^{-\frac{(9m+3)^2}{9^2\tau_2}}\equiv
\sum_{m\in\mathbb Z}e^{-\frac{(3m+1)^2}{3^2\tau_2}},
\eea
$\Lambda_1^{(3)}(\tau)\equiv\Lambda_3^{(9)}(\tau)$ and 
$\hat\Lambda_1^{(3)}(\tau)\equiv\hat\Lambda_3^{(9)}(\tau)$
and so there is a more interesting version of (\ref{zetanove}):
\bea
\label{zetanovebis}
Z^{(9)}=\int_{\mathcal S}  \frac{d^2\tau}{\tau_2^2}
\sum_{i=0}^8\frac{\ortw{0}{i}{9}}{{\Lambda_i}^{(9)}(\tau)}\times
\sum_{m\in\mathbb Z} e^{-\frac{(9m+i)^2}{9^2\tau_2}}.
\eea
It is easy to understand that formula (\ref{zetanovebis}) can be extended to 
every other case, independently of the form of the initial multiplet.
\bea
Z^{(N)}=\int_{\mathcal S}  \frac{d^2\tau}{\tau_2^2}
\sum_{i=0}^N\frac{\ortw{0}{i}{N}}{{\Lambda_i}^{(N)}(\tau)}\times
\sum_{m\in\mathbb Z} e^{-\frac{(Nm+i)^2}{N^2\tau_2}}.
\label{zetanoveend}
\eea

An important thing to note is that while the so-called ``first step'' is completely general, the 
``second step'' requires that we take a choice and pick a special object $\Lambda_i^{(n)}$.
This does not spoil the fact that the procedure can be applied to a generic multiplet, independently of 
its origin.
Nevertheless it is
interesting to note that strictly dependently on the details of the form of the various terms one can
pick a different and more suitable object to perform the second step.
In our case we can specialize (\ref{zetanoveend}) noting that, due to the translation action, a 
lattice term
\bea\nn
{\Lambda_i}^{(N)}(R,\tau)=\sum_{m,\,n\,\in \mathbb Z} e^{-\frac{R^2\left|N m+1+N n\tau\right|^2}{N^2  \tau_2}}
\eea
is present in $\ortw{0}{i}{N}$. We can use this
lattice to perform the second step, in such a way that the denominator is canceled by the 
lattice ${\Lambda_i}^{(N)}(R,\tau)$, obtaining the final result, valid for any odd $N$,
\be
\label{torozn}
Z^{(N)}=\frac{v_7}{N}\frac{R}{\sqrt{\al}}
\int_0^\infty\!\!\frac{d\tau_2}{2 \tau_2^5}\sum_{k=1}^{N-1}
\sum_{M\in\mathbb N}\left[ d_M(B)^k - d_M(F)^k \right] e^{-4\pi\tau_2 M}
\sum_{n\in\mathbb{Z}}
e^{-\frac{\pi R^2}{\alpha^\prime \tau_2}\left(n+\frac{k}{N}\right)^2},
\ee
where we have defined the coefficients $d_M(B,F)^k$ as
\be
\sum_{M\in\mathbb N}\left[ d_M(B)^k - d_M(F)^k \right] e^{-4\pi\tau_2 M}
=\int\!\! \frac{d\tau_1}{4}
\Bigg|
2\sin (\frac{2\pi k}{N})
\frac{
\sum_{\alpha,\beta}\eta_{\alpha\beta}
\theta{\alpha\brack \beta}^3
\theta{\alpha \brack \beta+\frac{2k}{N}}^3}
{\eta^9\,\,
\theta{\frac 12 \brack \frac 12+\frac{2k}{N}}}
\Bigg|^2\,,
\label{defdm}
\ee
and $v_7$ equals the ratio between the volume of the seven dimensional space time and a factor
$(4\pi^2\al)^{7/2}$.

The even-$N$ case can be easily deduced.
The multiplet is reduced in the same way, the only exception being that the term $\ortw{0}{N/2}{N}$ 
is special, because $\theta^{N/2}$ acts only as $(-1)^F$ and so while the fermionic part is as in 
(\ref{zetanoveend}) the bosonic is different and, in particular, there is an extra factor from
the momentum in the $\mathbb C$ dimensions.
So, to conclude, the general amplitude is described in (\ref{torozn})
with $k$ taking values $\{1,\dots,(N-2)/2,(N+2)/2,\dots,N-1\}$ and an extra term from the term $k=N/2$ of the
form
\bea
\label{exc}
\Delta=\frac{v_7 v_2}{N}\frac{R}{\sqrt{\al}}
\int_0^\infty\!\!\frac{d\tau_2}{2 \tau_2^6}
\sum_{M\in\mathbb N}\left[ d_M(B)^{N/2} - d_M(F)^{N/2} \right] e^{-4\pi\tau_2 M}
\sum_{n\in\mathbb{Z}}
e^{-\frac{\pi R^2}{\alpha^\prime \tau_2}\left(n+\frac{1}{2}\right)^2},
\eea
where
\bea
\sum_{M\in\mathbb N}
\left[ d_M(B)^{N/2} - d_M(F)^{N/2} \right] e^{-4\pi\tau_2 M}=
\int d\tau_1\left|\frac{\theta{1/2\brack 0}^4}{\eta^{12}}\right|^2
\eea
In (\ref{exc}) there is an extra volume $v_2$ taking in account the fact that
these states propagates also in the $\mathbb C$ dimensions.

The $\Z_4$ case is not special, and the formula described above is still valid.
The only remark we can make is the fact that starting from the two different elements 
we obtain the same result. This is true because the Scherk-Schwarz present in
$\ortw{0}{1}{4}$ case is that used usually, $\Lambda_1^{(4)}(R,\tau)$, while  $\ortw{2}{1}{4}$ 
there is a lattice that is 
exactly $ST^2S\{\Lambda_1^{(4)}(R,\tau)\}$ so that the unfolding for $\ortw{2}{1}{4}$ proceed as in 
$\ortw{0}{1}{4}$ but with an extra $ST^2S$ that transforms $\ortw{2}{1}{4}$ in $\ortw{0}{1}{4}$.

The amplitude given in (\ref{torozn}) has a clear field theory interpretation.
Given any field theory the one-loop vacuum energy can be written as
\bea\nn
E_D=-\int_0^\infty\frac{dt}{2t}Tr e^{-t H}
\eea
where the trace is performed over the full Hilbert space and $H=k^2+m^2$. If we have a 
theory with a spectrum containing $d_i$ particles of spin $J_i$ and mass $m_i$ 
propagating in $D$ dimensions, we can write, rescaling in a proper way $t$,
and considering an energy density rather then an energy
\bea\nn
\rho_D=-\int_0^\infty
\frac{dt}{2t^{\frac{D+2}{2}}}\sum_i d_i (-1)^{2 J_i} e^{-\pi t\al m_i^2}.
\eea
In our string case we have
\bea\nn
m^2_{i,n}=\frac{4 M}{\al}+\frac{(n+q)^2}{R^2}
\eea
where the term proportional to $M\in\mathbb N$ is the usual closed string mass and the second term
is the mass contribution from the Kaluza-Klein excitations $n\in\mathbb Z$, with the insertion of the 
twisted charge $q\in\{0,1/N,2/N,\dots\}$ given by the SS breaking, so that, for a model with states
propagating in $7$ extended plus one compact dimension with twists from $\Z_N$
\bea
\rho_7^{(N)}=-\int_0^\infty
\frac{dt}{2t^{\frac{9}{2}}}\sum_{M=0}^\infty\sum_{n\in\z}
\sum_{k=0}^{N-1}\sum_{F=0,\,1} d_M(k,F)(-1)^F e^{-4\pi M t}e^{-\frac{\pi t \al n^2}{R^2}}.
\label{ft}
\eea
that can be directly compared with the energy density that is derived from (\ref{torozn}) in the
odd-$N$ case. The even-$N$ case can be treated in the same way, but we must remember
that the particles of charge $N/2$ propagate in nine extended dimensions rather than in seven.
We take $\tau_2\rightarrow t$, $R\rightarrow RN$ and we perform a Poisson resummation in the
Kaluza-Klein lattice and (\ref{torozn}) can be written as
\bea
\rho_7^{(N)}=-\frac{R}{\sqrt{\al}}
\int_0^\infty\!\!\frac{d\tau_2}{2 \tau_2^5}\sum_{k=1}^{N-1}
\sum_{M\in\mathbb N}\left[ d_M(B)^k - d_M(F)^k \right] e^{-4\pi\tau_2 M}
\sum_{n\in\mathbb{Z}}
e^{-\frac{\pi R^2}{\alpha^\prime \tau_2}\left(Nn+k\right)^2}
\label{toroznft}
\eea
It is easy to see how after a Poisson resummation (\ref{toroznft}) reproduces (\ref{ft}) 
provided that 
\bea
\label{rel}
\frac{1}{N}\sum_{k=0}^{N-1} \left[d_M{B}^k-d_M(F)^k\right]e^{2\pi i\frac{k}{N}}P_0^N(n)=
\sum_{k=0}^{N-1}  \left[d_M(k,0)-d_M(k,1)\right]P_k^N(n)
\eea
where $P_k^N(n)$ equals 1 if $n=k\,\,mod\,\,N$, zero otherwise.
The relation (\ref{rel}) is not easily solved in the general case.
We solved it in the  $n=3$, $n=5$ cases checking that the coefficients $d_0(k,i)$ obtained from
(\ref{rel}) are the same we can obtain directly computing the
charges $q$ of the massless IIB string states.
In \cite{root} a detailed computation of this check is shown.

\begin{center}
\begin{figure}[t]
\hspace{.5cm}
\includegraphics[scale=0.65]{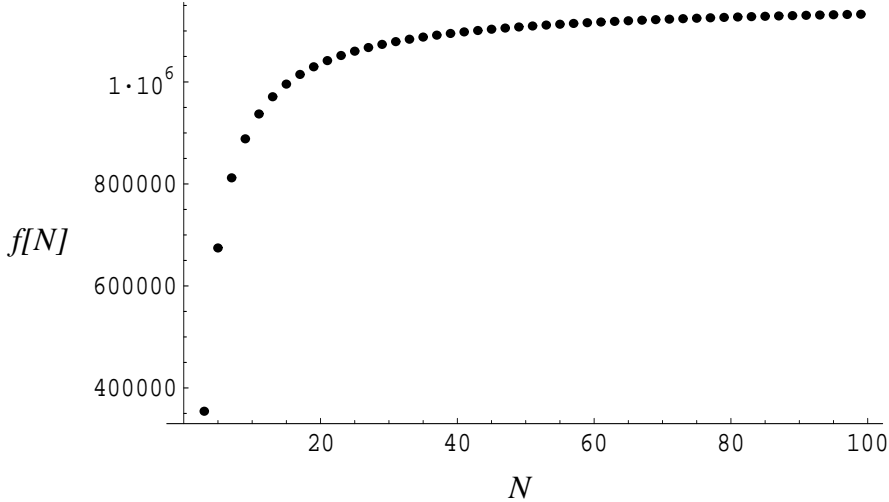}\vspace{-.7cm}
\caption{\small
The value of the $f[N]$ function in different cases. As clear $f[N]$ is a monotonic function
with asymptotic value $f[\infty]\sim 1.16*10^6$.
}
\label{Fig2}
\end{figure}
\end{center}

It is easy to compute the leading term for $Z^{(N)}$, given by the $M=0$ contribution.
It is given by
\bea\nn
Z^{(N)}\sim v_7\left(\frac{\al}{R^2}\right)^{\frac{7}{2}} \frac{3}{\pi^4}f[N]+
\frac{1+(-1)^N}{2} \frac{v_7v_2}{N}\left(\frac{\al}{R^2}\right)^{\frac{9}{2}}
\frac{c}{\pi^5}
\eea
where the function $f[N]$ contains all the $N$-dependence in the odd-$N$ case
and $c\sim 3\cdot 2^{22}$
\bea\nn
f[N]\sim\frac{2^9}{N}\sum_{k=1}^{(N-1)/2}
\frac{sin\left(\frac{\pi k}{N}\right)^8}{\left(\frac{k}{N}\right)^8}.
\eea
The behavior of $f[N]$ is summarized in fig.~\ref{Fig2}, it is essentially monotonic and in the limit
$N\rightarrow\infty$ it approaches the constant value
\bea\nn
Lim_{N\rightarrow\infty} f[N] =
f[\infty]=2^9\int_0^1 dx \left(\frac{sin(\pi x)}{x}\right)^8\sim 1.16*10^6.
\eea
The monotonic behavior ensures that the one parameter set of functions 
$Z^{(N)}$ has, at least at the leading order, a monotonic behavior in $N$
in the odd-$N$ case, as found in the lowest $N$ case
in \cite{root}.
In the even-$N$ case the extra factor from the term $k=N/2$ changes the 
behaviour, that depends also on the ratio between the volume $v_2$ and $R^2$.

\noindent
{\Large \bf Acknowledgments} 
\vskip 10pt

\noindent
I would like to thank Marco Serone for his guidance during the 
preparation of this work. 
I would like also to thank Monica Borunda for
useful discussions in the early stage of the work.
This work was partially supported by the EC through the RTN network ``The quantum structure 
of space-time and the  geometric nature of fundamental interactions'', contract HPRN-CT-2000-00131.

\appendix
\section*{Appendix}
In this appendix we describe the one loop closed string amplitude (partition function) for a 
generic $\Z_N$ orbifold of type IIB string theory. 
The main task is to obtain the form of the $\ortw{i}{j}{N}$ 
objects used in the paper an their modular properties, in particular we give a 
recipe to study the net of identifications between the $\ortw{i}{j}{N}$ under a modular transformation.
We make a large use of the general formulas derived in \cite{SeSc}.

A IIB orbifold model is given when we consider a non trivial background
$\mathbb R^4\times M_6$ for type IIB string theory, where $M_6$ is a six dimensional manifold of the form
$\mathbb C^n\times T^{3-n}/\Z_N$, $T$ being a two-dimensional torus and $n\in\{0,1,2,3\}$.
The value of $n$ is relevant only for the form of the bosonic zero-modes, the 
momentum of the string.
From the point of view of the modular transformation this is not relevant.

We take $\Z_N$ generated by an order $N$ rotation in the various $\mathbb C$ or $T$
manifold, the rotation angle being $2\pi\times\vec{v}$. The vector $v$ is of the form
\bea\nn
\vec{v}=\frac{1}{N}(a,b,c)
\eea
with $a,b,c$ natural numbers. World-sheet supersymmetry impose that the rotation is applied also
on world sheet fermions.  In order to have that the action of the rotation is of order $N$ 
also on the fermionic degrees of freedom  we require that $a+b+c$ is an even number.
We do not impose any other constraint since we are interested also in non-SUSY orbifolds.

The partition function is given by
\bea\nn
Z=\frac{1}{N}\int_{\mathcal F}\frac{d^2\tau}{2\tau_2^2}\sum_{i,\,j\,=0}^{N} \ortw{i}{j}{N}
\eea
where
\bea\nn
\ortw{i}{j}{N}=
Tr\Bigg|_i\left[\theta^j q^{L_0}\bar q^{\bar L_0}\right],\,\,\, q=e^{2 \pi i \tau}
\eea
where $\theta$ is the $\Z_N$ generator and the trace is performed over the light-cone degrees of
freedom of the $\theta^i$ twisted sector, i.e. the sector where the world-sheet fields are identified as
\bea\nn
\Phi(s+2\pi,t)=\theta^i \Phi(s,t),
\eea
$s$ and $t$ being the world-sheet coordinates.
Clearly in the trace also a GSO projection and a sum over the spin structures is intended.

We can also join the described rotation with a translation in some compact direction where 
the rotation acts in a suitable way, obtaining a so-called freely acting orbifold.
It is not necessary to impose any requirements about the relation between rotation and translation
at this stage. 
We describe a completely general case giving the building blocks for $\ortw{i}{j}{N}$, without
any constraint. Clearly, when putting together the terms to build the partition function, the 
requirements must be fulfilled.

Since in the theory the fermionic modes are decoupled from the bosonic ones and, furthermore,
the bosonic zero modes are decoupled from the other bosonic modes, $\ortw{i}{j}{N}$ factorize 
in the product of three
terms: a term from bosonic non-zero modes, one from fermionic modes 
and one from momentum, that is the bosonic zero mode. This last 
term contains also a factor arising
from the number of fixed points of the operators $\theta^i$ 
and  $\theta^j$, as described in the appendix of \cite{sst}.

The fermionic modes give the contribution
\bea
\pportw{h}{k}{N}_F\hspace{-10pt}(\tau) &=& 
\left|
\sum_{\alpha,\,\beta\in\{0,1/2\}}\frac{(-1)^{2\alpha+2\beta+4\alpha\beta}}{2}
\prod_{d=0}^3\eta^{-1}(\tau)\,e^{-2 \pi h \beta v_d}\,
\theta{\alpha+h v_d \brack \beta+k v_d}(\tau)
\right|^2
\label{ZF}
\eea
where $v_0$ is usually taken to be zero while $v_1$, $v_2$ and $v_3$ 
are the components of the above described 
$\vec{v}$. 
The $\theta$ functions are the well-known modular functions whose 
definition and modular properties are described,
for example, in the appendix of \cite{SeSc}.

The sum over the spin structure can be simplified using the Jacobi 
identity described in the appendix A of 
\cite{kir}, using it (\ref{ZF}) is written as
\bea
\pportw{h}{k}{N}_F\hspace{-10pt}(\tau) &=& 
\left|
\prod_{d=0}^3\eta^{-1}(\tau)\,
\theta{1/2+h v^\prime_d \brack 1/2+k v^\prime_d}(\tau)
\right|^2
\label{ZF2}
\eea
where 
\bea
v_0^\prime=\frac{1}{2}(v_1+v_2+v_3),& v_1^\prime=\frac{1}{2}(-v_1+v_2+v_3)\nn\\
v_2^\prime=\frac{1}{2}(v_1-v_2+v_3),& v_3^\prime=\frac{1}{2}(v_1+v_2-v_3).\nn
\eea
From the modular properties of the $\theta$ functions it is easy to see that
$T$ maps $\pportw{i}{j}{N}_F\hspace{0pt}(\tau)$ in 
$\pportw{i}{i+j}{N}_F\hspace{0pt}(\tau)$, while $S$ maps $\pportw{i}{j}{N}_F\hspace{0pt}(\tau)$ in 
$\pportw{j}{-i}{N}_F\hspace{0pt}(\tau)$. It is also interesting to note that, due to the periodicity properties,
$\pportw{i}{j}{N}_F\hspace{0pt}(\tau)\sim\pportw{-i}{-j}{N}_F\hspace{0pt}(\tau)$ 
and $\pportw{i}{j}{N}_F\hspace{0pt}(\tau)\sim\pportw{aN+i}{bN+j}{N}_F\hspace{0pt}(\tau)$ for
all $a,\,b\in\mathbb N$.

The term from the bosonic non-zero modes can be split in the product of four terms related to the four $v_i$.
Each term is of the form
\bea\nn
\pportw{h}{k}{N}_{B}^{(d)}\hspace{-10pt}(\tau)=
\left|\eta(\tau)\right|^{-4}
\eea
if $h v_d,\,k v_d\in\mathbb{N}$
or of the form
\bea\nn
\pportw{h}{k}{N}_{B}^{(d)}\hspace{-10pt}(\tau)=
\Bigg|\eta(\tau) \, 
\theta^{-1}{\frac 12 + h v_d \brack \frac 12 + k v_d}(\tau)\Bigg|^2
\eea
in the other cases.

The term from the bosonic zero modes, like the previous one, can be split in the product of four terms,
each of them having a different form depending on the action of a translation/rotation.
Each term equals one if a non-trivial rotation is present, 
if no rotation is present it equals
\bea
\pportw{h}{k}{N}_{0B}^{(d)}\hspace{-10pt}(\tau)=
\frac{\sqrt{G}}{\alpha^\prime\,\tau_2} \,
\sum_{\vec m, \vec n} e^{- \frac {\pi }{\alpha^\prime\tau_2} 
[(m + k w_d) + (n + h w_d)\tau]_i (G+B)_{ij} 
[(m + k w_d) + (n + h w_d) \bar \tau]_j}
\label{latt}
\eea
if the submanifold is compact and the $\Z_N$ acts also as a translation of length $\hat w_d$ in this direction.
The two dimensional lattice is written in the most general case, with $G$ and $B$ respectively the metric and
antisymmetric field. Also the translation is generic, we only ask that it is of order $N$, without any other 
constraint on the form of the two-dimensional vector $w_d$.
The formula (\ref{latt}) is extended in the non-compact case and resummed to $V/4\pi^2\al\tau_2$, $V$ being, as
usual, the infinite  volume of the two-dimensional non-compact manifold itself.
Clearly in presence of a non compact manifold it is meaningless to introduce an order $N$ translation.

Furthermore there is an overall factor $\pportw{h}{k}{N}_{\mathcal N}$ counting the number of fixed points.

The full $\ortw{h}{k}{N}$ is given by
\bea
\pportw{h}{k}{N}(\tau)=
\pportw{h}{k}{N}_{\mathcal N}
\pportw{h}{k}{N}_F\hspace{-10pt}(\tau)
\prod_{d=0}^4\pportw{h}{k}{N}_B^{(d)}\hspace{-10pt}(\tau)\pportw{h}{k}{N}_{0B}^{(d)}\hspace{-10pt}(\tau)
\eea
and it has the same modular and periodicity properties of $\pportw{h}{k}{N}_F\hspace{0pt}(\tau)$, so that
\bea
\bcd
\ortw{h}{k}{N} @>S>>    \ortw{k}{-h}{N} \\
       @VTVV              @.      \\
\ortw{h}{h+k}{N}  @.\\   
\ecd
\eea
and
$
\ortw{h}{k}{N}\sim\ortw{-h}{-k}{N}\sim\ortw{nN-h}{pN-k}{N},\,\,\forall\,\,n,\,p\,\in\mathbb N.
$


\begin{thebibliography}{99}

\bibitem{Tan}
K.~H.~O'Brien and C.~I.~Tan,
Phys.\ Rev.\ D {\bf 36} (1987) 1184.

\bibitem{Dix}
L.~J.~Dixon, V.~Kaplunovsky and J.~Louis,
Nucl.\ Phys.\ B {\bf 355} (1991) 649.
%
\bibitem{MS}
P.~Mayr and S.~Stieberger,
Nucl.\ Phys.\ B {\bf 407} (1993) 725
[arXiv:hep-th/9303017].
%
\bibitem{Mar}
M.~Marino and G.~W.~Moore,
Nucl.\ Phys.\ B {\bf 543} (1999) 592
[arXiv:hep-th/9808131].
%
\bibitem{Ghil}
D.~M.~Ghilencea, H.~P.~Nilles and S.~Stieberger,
arXiv:hep-th/0108183.

\bibitem{SS}
J.~Scherk and J.~H.~Schwarz,
Phys.\ Lett.\ B {\bf 82} (1979) 60;
Nucl.\ Phys.\ B {\bf 153} (1979) 61.

\bibitem{ads1}
I.~Antoniadis, E.~Dudas and A.~Sagnotti,
Nucl.\ Phys.\ B {\bf 544} (1999) 469
[arXiv:hep-th/9807011].
%
\bibitem{adds}
I.~Antoniadis, G.~D'Appollonio, E.~Dudas and A.~Sagnotti,
Nucl.\ Phys.\ B {\bf 553} (1999) 133
[hep-th/9812118];
Nucl.\ Phys.\ B {\bf 565} (2000) 123
[hep-th/9907184]; \\
I.~Antoniadis, K.~Benakli and A.~Laugier,
[hep-th/0111209].
%
\bibitem{SeSc}
C.~A.~Scrucca and M.~Serone,
JHEP {\bf 0110} (2001) 017
[arXiv:hep-th/0107159].
%
\bibitem{sst}
C.~A.~Scrucca, M.~Serone and M.~Trapletti,
Nucl.\ Phys.\ B {\bf 635} (2002) 33
[arXiv:hep-th/0203190].
%
\bibitem{Cot}
A.~L.~Cotrone,
Mod.\ Phys.\ Lett.\ A {\bf 14} (1999) 2487
[arXiv:hep-th/9909116].

\bibitem{BD}
J.~D.~Blum and K.~R.~Dienes,
Nucl.\ Phys.\ B {\bf 516} (1998) 83
[arXiv:hep-th/9707160].

\bibitem{root}
M.~Borunda, M.~Serone and M.~Trapletti,
arXiv:hep-th/0210075.

\bibitem{kir}
E.~Kiritsis,
arXiv:hep-th/9709062.

\bibitem{Pol}
J.~Polchinski,
``String Theory. Vol. 1: An Introduction To The Bosonic String,''
{\it  Cambridge, UK: Univ. Pr. (1998) 402 p}.

\end{thebibliography}
\end{document}